\documentclass[11pt]{article}
\usepackage{epsfig}
\usepackage{latexsym}
\usepackage{amsmath}

\newtheorem{theorem}{Theorem}[section]{\bf}{\it}
{\bf}{\rm}
\newtheorem{proposition}[theorem]{Proposition}{\bf}{\it}
\newtheorem{corollary}[theorem]{Corollary}{\bf}{\it}
\newtheorem{lemma}[theorem]{Lemma}{\bf}{\it}
\renewcommand{\forall}{\mbox{for all}\,\,}
          \def\dt{\cal}
          
          \def\dB{{\dt B}}

          \def\dM{{\dt M}}
          \def\dN{{\dt N}}

          \def\E{{\cal E}}

          \def\H{{\cal H}}

          \def\O{{\cal O}}
          \def\P{{\cal P}}

      \def\gO{\Omega}
          \def\go{\omega}

          \def\complex{{\bf C}}
          
      \def\ddt{\frac{{\rm d}}{{\rm dt}}}
          
      \def\eps{\varepsilon}

          \def\h{{\bf h}}
          \def\Halmos{\quad\hfill$\Box$}
          
          \def\Im{{\rm Im}\,}
          \def\impuls{k}

          \def\modop{\Delta^{1/2}}
          
          \def\modopt{\Delta^{it}}
          \def\modopmt{\Delta^{-it}}
          \def\naturals{{\bf N}}
          \def\pct{P$_1$CT}

          \def\Rd{\reals^{1+s}}
          \def\reals{{\bf R}}

          \def\tr{{\rm tr}}
\title{Covariant Thermodynamics of Quantum Systems:
Passivity, Semipassivity, and the Unruh Effect}
\author{Bernd Kuckert\\University of Amsterdam\\
Korteweg-de Vries Institute for Mathematics\\Plantage Muidergracht 24\\
1018 TV Amsterdam, The Netherlands\\e-mail: kuckert@science.uva.nl}
\date{November 2001}

\begin{document}
\maketitle

\begin{abstract}
According to the Second Law of Thermodynamics, cycles applied to
thermodynamic equilibrium states cannot perform any work (passivity
property of thermodynamic equilibrium states). In the presence of matter
this can hold only in the rest frame of the matter,
as moving matter drives, e.g., windmills and turbines.
If, however, a homogeneous and stationary state has the property that no cycle
can perform more work than an ideal windmill, then it can be shown
that there is some inertial frame where the state is a
thermodynamic equilibrium state.
This provides a covariant characterization of thermodynamic equilibrium
states.

In the absence of matter, cycles should perform work only when driven by
nonstationary inertial forces caused by the observer's motion.
If a (pure) state of a relativistic quantum field theory behaves
this way, it satisfies the spectrum condition and exhibits
the Unruh effect.

\end{abstract}

\section{Introduction}

In \cite{PW78}, Pusz and Woronowicz analyzed thermodynamic equilibrium
states in a general quantum theoretical setting. They established
that the condition of complete passivity, which can be derived from
the first principles of thermodynamics, is, at nonzero temperature,
equivalent to the
Kubo-Martin-Schwinger (KMS-) condition, a widely used
generalization of the Gibbs characterization of thermodynamic
equilibrium states that is appropriate also for systems with
infinitely many degrees of freedom \cite{Kubo, MarS,HHW,Haa92}.

Complete passivity is a strengthening of the principle that
cycles, when applied to thermodynamic equilibrium
states, cannot perform any work, which is the passivity property of
thermodynamic equilibrium states. In the
presence of matter, however, a system can exhibit passivity
in one distinguished
frame of reference at most, as moving matter can drive cycles
(e.g., windmills and turbines). So the question arises
what a covariant version of the (complete) passivity condition
could look like.
This problem is of particular interest if the state under consideration
is stationary and homogeneous, as its invariance properties alone do not
distinguish any frame of reference in this case.
If a thermodynamic system is
covariant under a representation of the spacetime translation group, it
possesses at least one stationary and homogeneous nonequilibrium state
\cite{Nar77}, and thermal equilibrium states in quantum field
theories exhibit themselves as nonequilibrium states to
moving observers \cite{Nar77,Oji86}.

In this article, the Pusz-Wornowicz analysis
will be generalized to a version that can be used in each
inertial frame of reference. It is well known that
the power of a windmill or a turbine depends on the third power of the
velocity of the current driving the device. It will be shown
in Sect. \ref{covariant version} that if a cycle applied
to a stationary and homogeneous state $\go$ cannot perform more
work than such a device, then
there is a frame of reference where the considered state is a thermodynamic
equilibrium state in the sense that it satisfies the KMS-condition or is a
ground state of the Hamiltonian,
which corresponds to the case of zero temperature.

The first covariant characterization of
thermal states in quantum field theory was recently given by Bros and
Buchholz \cite{BB94}.
Their criterion is a relativistic KMS-condition, and they expected that this
condition could eventually be derived from the assumption that the
work a cycle can extract from the system within a given spacetime
region is bounded by a constant that depends on the size of the
spacetime region. This conjecture motivated the analysis to follow
and is partially confirmed by its results.

In Sect. \ref{chemical potential} it is briefly discussed
how the results of Sect. \ref{covariant version} can be applied
to describe the chemical potential.

In Sect. \ref{vacuum}, the notion of a vacuum state is discussed.
Due to the
absence of matter and energy flows, only nonstationary inertial forces
caused by an acceleration of the observer should drive cycles when
applied to a vacuum state. It is shown that the spectrum condition
holds for each pure state behaving this way.

In Sect. \ref{BW-theorem}, it is shown that
if a state of a local quantum field theory behaves this way, then
it exhibits the Unruh effect. The Unruh effect first has been
established by Unruh \cite{Unr76} for the free field and,
independently and simultaneously,
by Bisognano and Wichmann \cite{BW75,BW76} for finite-component
Wightman fields. A recent derivation for
massive particle states in algebraic quantum field theory is due to
Mund \cite{Mun01}, and a couple of uniqueness results concerning
the Unruh effect in this setting can be found in \cite{Bor00,Kuc00}
and the references quoted there.

Results similar to those of Sects. \ref{vacuum} and \ref{BW-theorem}
have recently been presented in \cite{BB99,BFS99} for quantum
fields on de Sitter and Anti-de Sitter spacetimes,
respectively: assuming that a given state exhibits
the KMS-condition to all uniformly accelerated observers,
covariance under a representation
of the spacetime's symmetry group was established, and the
corresponding Hawking temperatures were computed.

Sect. \ref{conclusion} summarizes the results.

\section{Thermodynamic equilibrium and passivity}
\label{intro}

In the quantum statistical mechanics of general quantum systems
\cite{BR1,BR2}, each quantum system is characterized by its
algebra $\dM$ of observables, which we here assume to be a von Neumann
algebra\footnote{A von Neumann algebra is a linear space
$\dM$ of bounded operators in $\H$ that contains the adjoint of each
of its elements and the operator product of any two of its elements,
and that coincides with its {\it bicommutant} $\dM''$; here
$\dM'$ denotes the {\it
commutant} of $\dM$, i.e., the (von Neumann) algebra of all bounded
linear operators that commute with all elements of $\dM$, and
$\dM'':=(\dM')'$. One could choose $\dM$ within a larger class of
algebras as well, namely, the C$^*$-algebras with a unit element, but
to save notation, we confine ourselves to von Neumann algebras.}.
Each state of the system is described by a
linear functional $\go$ on $\dM$
that associates with each $A\in\dM$ the corresponding
expectation value $\go(A)\in\complex$ such that
$\go(A^*A)\geq0$ for all $A\in\dM$ and $\go(1)=1$.

In what follows, one state $\go$ will be our object of
investigation, and it will be assumed that this state
is induced by a cyclic vector $\gO$ of $\dM$
(i.e., $\go(A)=\langle\gO,A\gO\rangle$, and
$\overline{\dM\gO}=\H$). The existence of $\gO$
can be assumed without
loss of generality by using the GNS-representation
of $\go$, as all properties
of $\dM$ assumed below are inherited by this representation.

A selfadjoint operator $H$ will be considered as the free
Hamiltonian of the system: for each $A\in\dM$ and each
time $t\in\reals$,
it is assumed that $A_t:=e^{itH}Ae^{-itH}$ lies in $\dM$.
We consider the case that $\go$ is stationary with respect to this
time evolution, i.e., that $\go(A)=\go(A_t)$ for all $A\in\dM$
and all $t\in\reals$. In this case, $\gO$ is an eigenvector of $H$, and
subtracting the corresponding eigenvalue from $H$, one can choose
$H$ such that $H\gO=0$.

If $e^{-\beta H}$ is trace class for a $\beta\geq0$, then
$\go$ is a thermodynamic equilibrium state if
it is a Gibbs state at the inverse temperature $\beta$.
In this case,
the two-point
function $z\mapsto(\tr(e^{-\beta H}))^{-1}\tr(e^{-\beta H}\,
e^{izH}Be^{-izH}A)=:F(z)$ is
analytic in the open strip
$S_\beta:=\{z\in\complex:\,-\beta<\Im z<0\}$
and continuous on $\overline{S}_\beta$, and it
satisfies
\begin{equation}\label{KMS}
F(t)=\go(B_t A)\quad\mbox{and}\quad
F(t-i\beta)=\go(AB_t)\quad\forall t\in\reals,\quad A,B\in\dM.
\end{equation}
In general, $\go$ is called a
{\it KMS-state (at the inverse temperature $\beta$)} of the dynamics generated
by $H$ if there exists a continuous function $F:\overline{S}_\beta\to\complex$
that is analytic in $S_\beta$ and satisfies the boundary condition (\ref{KMS}).
While the KMS-condition is equivalent to the Gibbs condition for
finite volume systems, it remains a
meaningful condition in the general
case (of, e.g., an infinitely extended system)
that $e^{-\beta H}$ is not trace class,
as it merely refers to the two-point function of $\go$.
For this reason, the KMS-condition is used to characterize
thermodynamic equilibrium states in such cases (\cite{HHW},
cf. also \cite{Haa92}). KMS-states at infinite temperature
are traces, i.e., states satisfying $\go(AB)=\go(BA)$ for all
$A,B\in\dM$, and
$\go$ may be considered a ``KMS-state at zero temperature''
if $H\geq0$, i.e., if $\go$ is a {\it ground state} of $H$.

Physically, a thermodynamic equilibrium state can be
characterized by its reaction to a cyclic reversible change of
the external conditions, a {\em cycle}.
For the time being, we consider as
a cycle any perturbation of
$H$ by (time-dependent) self-adjoint elements $h(t)$ of $\dM$
that depend on $t\in\reals$ in a norm-continuously differentiable
fashion and vanish for $t\notin[0,T]$
for some $T>0$. The duration $t_h$ of the cycle $h$ is the
smallest $T>0$ satisfying this condition.

Dyson's perturbation series yields the perturbed unitary time evolution
\newline $(U_h(t))_{t\geq0}$ that solves the equation
\begin{equation}\label{Dyson}
\ddt U_h(t)=-ie^{itH}h(t)e^{-itH}U_h(t),\qquad t\geq0,
\end{equation}
with the initial condition $U_h(0)=1$.
If $\gO$ is the system's state vector at $t=0$, this time
evolution determines the state vector $\psi(t):=e^{-itH}U_h(t)\gO$
for $t>0$;
formally, this is expressed by the Schr\"odinger equation
$i\ddt\psi(t)=(H+h(t))\psi(t)$. The expectation value of the
rate at which the process $h$ adds
energy to the system at the time $t$
is $\langle\psi(t),\dot{h}(t)\psi(t)\rangle$, and
the expectation value of the energy added
to the system by the time $t_h$ is
$$L_h:=\int_0^{t_h}\langle\psi(t),
\dot{h}(t)\psi(t)\rangle\,dt.$$
As $h$ is a reversible process,
there is no heat extracted from or added to the
system, so by the First Law of Thermodynamics,
$L_h$ is the work required for the cycle $h$;
equivalently, $-L_h$ is the work performed
by $h$. The Second Law of Thermodynamics requires that
$-L_h\leq0$ for all cycles $h$ if $\go$ is a
thermodynamic equilibrium state.

$\go$ is called a {\em passive} state if $-L_h\leq0$
for all cycles $h$. Let $U(\dM)$ denote the group of unitary
elements of $\dM$ and $U_1(\dM)$ the norm-connected component
of $U(\dM)$ that contains the unit operator.
By Thm. 2.1 in \cite{PW78},
$\go$ exhibits passivity if and only if one has
\begin{equation}\label{u-passivity}
-\langle W\gO,[H,W]\gO\rangle=-\langle W\gO,HW\gO\rangle\leq0
\end{equation}
for all $W\in U_1(\dM)$ with $[H,W]\in\dM$.\footnote{The
expression $\langle Hx,Wy\rangle-\langle x,WHy\rangle$
is defined for all $x,y$ in the domain of
$H$. $[H,W]\in\dM$ means that the sesquilinear form defined this way
is bounded and that the associated bounded operator is an element of
$\dM$; if commutators involving one unbounded selfadjoint operator are
referred to as elements of $\dM$ in what follows, this is to be
read this way. Note that the vector $A\gO$ is in the domain
of $H$ for all $A\in\dM$ with $[H,A]\in\dM$ (cf. Prop. 3.2.55
in \cite{BR1} and p. 280 in \cite{PW78}).}
For the typical finite-temperature case that
$\go$ is known not to be a trace, a glance at the proof of this
theorem in \cite{PW78} shows that
passivity holds if and only if Ineq. (\ref{u-passivity}) holds for
{\it all} unitary elements $W$ of $\dM$ with $[H,W]\in\dM$.

The unitaries $W$ can be interpreted as
propagators $U_h(t_h)$: if $h$ is a cycle with $[H,h(t)]\in\dM$
for all $t\in\reals$, then
$L_h=\langle U_h(t_h)\gO,HU_h(t_h)\gO\rangle.$
In the form of Ineq. (\ref{u-passivity}),
the passivity condition does no longer depend
on the technical condition that
$h(t)$ depends on $t$ in a norm continuously differentiable
fashion; it applies to any cycle that provides some appropriate
unitary propagator $W\in\dM$ with $[H,W]\in\dM$.

While a mixture of passive states is passive,
a mixture of states at different temperatures
does not have any well-defined temperature and, hence, cannot be a
thermodynamic equilibrium state according to the Zeroth Law. It follows
that the class of passive states is considerably larger than that of
thermodynamic equilibrium states, and that a stronger assumption is
needed to single out the thermodynamic equilibrium states.

As one way to strengthen the notion of passivity accordingly,
the condition of complete passivity was used in \cite{PW78}.\footnote{
As an alternative, Pusz and Woronowicz strengthen the passivity
assumption by assuming a cluster property in addition, which
characterizes pure thermodynamic phases. It is
straightforward to check that our results to follow can be
modified accordingly.} It can be
physically motivated as follows. The Zeroth Law of thermodynamics requires
that if a system in thermodynamic equilibrium is coupled to
an identical system in thermodynamic equilibrium and at the
same temperature, the combined system should be in thermodynamic
equilibrium as well, and this should also hold if more than two
identical copies of the system are coupled. So if $\go$ is a
thermodynamic equilibrium state and if
$\dM^{\bigotimes N}$ is the $N$th tensorial power of $\dM$,
the product state defined by
$$\dM^{\bigotimes N}\ni A_1\otimes A_2\otimes\dots
\otimes A_N\mapsto\go(A_1)\go(A_2)\dots\go(A_n)$$
should be passive as well. $\go$ is then called
{\it completely passive}. As shown
in \cite{PW78}, a state is completely passive if and only if it is
a KMS-state or a ground state.

\section{Moving systems and semipassivity}\label{covariant version}

A more general version of (complete) passivity can be used to characterize
thermodynamic equilibrium states in a covariant fashion, i.e., by
a criterion that does not only hold in the system's rest frame.
This issue is of
interest if the state $\go$ is stationary and homogeneous: if $\go$ is
stationary in no inertial frame, there is no thermodynamic equilibrium
to characterize, and if it is stationary in a unique inertial frame,
this frame is already distinguished by this fact, and it suffices to
apply the results of Pusz and Woronowicz. But
stationarity with respect to two distinct time evolutions implies
invariance of $\go$ under translations in at least one spatial direction.

In what follows, we
consider the case that there are $s\geq1$ spatial directions with this
property. The above generator $H$
of the time translations and $s$ other self-adjoint operators
$P_1,\dots,P_s$, which we consider as the
generators of the spatial translations and which
we collect in the vector operator ${\bf P}$, will be assumed to
generate a unitary
representation $V$ of the (1+s)-dimensional translation group
$(\Rd,+)$ such that $V(x)AV(x)^*\in\dM$ for all $A\in\dM$ and all
$x\in\Rd$. $\go$ is assumed to be invariant under $V$, i.e.,
$\go(V(x)AV(x)^*)=\go(A)$ for all $A\in\dM$ and all
$x\in\Rd$, so $\gO$ is an eigenvector of ${\bf P}$, and
again, we can (without loss) choose ${\bf P}$ such that
$P_1\gO=\dots=P_s\gO=0$.

Suppose that $\go$ is passive with respect to the Hamiltonian $H$.
If the system is not at rest, but moves at a velocity ${\bf u}$,
then the time evolution is not generated by $H$, but by
$H+{\bf u P}$.\footnote{In a relativistic theory, this
generator must be multiplied by the time dilation factor
$\gamma=(1-{\bf u}^2/c^2)^{-1/2}$; see below.}
If $W\in U_1(\dM)$ satisfies $[H,W]\in\dM$ and $[{\bf P},W]\in\dM$,
then the work performed by the corresponding cycle is
\begin{align*}
-L&=-\langle W\gO,
(H+{\bf u}{\bf P})\,W\gO\rangle\\
&\leq-\langle W\gO,{\bf u}{\bf P}\,W\gO\rangle,
\end{align*}
as $\go$ is passive with respect to $H$.
Defining $|{\bf P}|:=\sqrt{P_1^2+\dots+P_s^2}$, one finds
$-{\bf u}{\bf P}\leq|{\bf u}|\,|{\bf P}|$,
so
\begin{equation}\label{presemip}
-L\leq |{\bf u}|\langle W\gO,|{\bf P}|\,W\gO\rangle.
\end{equation}

Now suppose that $\go$ is not necessarily passive with respect
to $H$. We call
$\go$ {\em semipassive} if the work a cycle can perform is
bounded as in Ineq. (\ref{presemip}), i.e., if
there is a constant $\E\geq0$ such that
\begin{equation}\label{semipassivity}
-\langle W\gO,H\,W\gO\rangle
\leq\E\langle W\gO,|{\bf P}|\,W\gO\rangle
\end{equation}
for all $W\in U_1(\dM)$ with
$[H,W]\in\dM$ and $[{\bf P},W]\in\dM$. The constant $\E$
will be referred to as an {\em efficiency bound} of $\go$.
Generalizing also the notion of complete passivity, $\go$
will be called a {\em completely semipassive}
state if all its finite tensorial powers are
semipassive with respect to one fixed efficiency bound $\E$.

By the above considerations, a state is completely
semipassive in all inertial frames if it is completely passive
in some inertial frame. We will prove now that if, conversely,
a state is completely semipassive in a given inertial frame, then there
exists an inertial frame where it is completely passive.

We proceed in
two steps by distinguishing the cases that $\go$ is faithful,
i.e., that given any $A\in\dM$, $\go(A^*A)=0$ implies $A=0$,
and that $\go$ is not faithful.
\begin{proposition}\label{rkms theorem}
Let $\go$ be completely semipassive with efficiency bound $\E$.
If $\go$ is faithful, then there exists a ${\bf u}\in\reals^s$
with $|{\bf u}|\leq\E$ such that

(i) $H+{\bf uP}=0$, or

(ii) $\go$ is a
KMS-state at finite $\beta\geq0$
with respect to $H+{\bf u}{\bf P}$.
\end{proposition}
{\it Proof.}
Using the reasoning that lead to Ineq.
(3.8) in \cite{PW78}, one can derive from semipassivity that the
useful inequality
\begin{equation}\label{windmill}
\langle A\gO,(H+\E|{\bf P}|) e^{-H^2-|{\bf P}|^2}A\gO\rangle
+\langle A^*\gO,(H+\E|{\bf P}|) e^{-H^2-|{\bf P}|^2}A^*\gO\rangle
\geq 0
\end{equation}
holds for {\it all} $A\in\dM$. The
factor $e^{-H^2-|{\bf P}|^2}$ is a mollifier that avoids domain problems.

As $\go$ is faithful, $\gO$ is separating, so Tomita-Takesaki theory
(cf. App. \ref{TTTheory}) defines the modular operator $\Delta$
and the infinitesimal generator $K=\ln(\Delta)$ of the modular automorphism
group. As the dynamics generated by $K$ satisfies the KMS-condition
and is, up to a scalar multiplication of $K$, the only such dynamics,
we compare $H$ and ${\bf P}$ with $K$.

As $H$ and ${\bf P}$ generate one-parameter unitary groups that act as
automorphisms on $\dM$ and leave $\gO$ fixed, they strongly commute
with $K$ (cf. App. \ref{TTTheory}).
One obtains from Ineq. (\ref{windmill}) (see \cite{PW78} for details) that
$$-H(1-\Delta)e^{-H^2-|{\bf P}|^2}
\leq\E|{\bf P}|(1+\Delta)e^{-H^2-|{\bf P}|^2}.$$
By this inequality, the joint spectrum\footnote{See App.
\ref{jspec} for some remarks on joint spectra.}
$\sigma_{H,{\bf P},K}$ of $H$, ${\bf P}$, and $K$ is a subset of
the set
$$\sigma^{(\E)}:=\{(\eta,{\bf\impuls},\kappa)\in\reals^{s+2}:\,
-\eta (1-e^\kappa)\leq\E|{\bf\impuls}|(1+e^\kappa)\},$$
which contains the entire $\kappa$=0-plane and all
$(\eta,{\bf k},\kappa)\in\reals^{s+2}$
with $\kappa>0$ and
\begin{equation}\label{constraint1}
\eta\leq\E|{\bf\impuls}|\frac{e^\kappa+1}{e^\kappa-1},
\end{equation}
and all $(\eta,{\bf k},\kappa)\in\reals^{s+2}$ with $\kappa<0$ and
\begin{equation}\label{constraint2}
\eta\geq-\E|{\bf\impuls}|\frac{1+e^\kappa}{1-e^\kappa}.
\end{equation}
Using complete semipassivity
and the identities $JHJ=-H$, $J{\bf P}J=-{\bf P}$, and $JKJ=-K$,
(cf. App. \ref{TTTheory})
one can argue like in \cite{PW78} to show that
$\sigma_{H,{\bf P},K}$
is a subset of a subgroup $\tilde\sigma_{H,{\bf P},K}$
of $(\reals^{s+2},+)$ that is a subset of
$\sigma^{(\E)}$, so the above estimates imply that it must be a
subset of an at most (s+1)-dimensional
subspace of $\reals^{s+2}$. The smallest such subspace $X$ must be
a subset of $\sigma^{(\E)}$ as well.
Namely, if $(\eta,{\bf k},\kappa)\in\sigma^{(\E)}$, then
$(\lambda\eta,\lambda{\bf k},\lambda\kappa)\in\sigma^{(\E)}$
for all $\lambda\in[0,1]$, so $\bigcup_{\lambda\in[0,1]}\lambda
\tilde\sigma_{H,{\bf P},K}\subset\sigma^{(\E)}$. As $X$ is
the closure of the left-hand side, and as $\sigma^{(\E)}$ is a
closed set, it follows that
$X\subset\sigma^{(\E)}$, as stated.

Alternative (i) states that $H$ is a linear function of ${\bf P}$.
In particular, this holds if $X$ contains the $\kappa$-axis:
by Lemma \ref{projection}, the joint spectrum of $H$ and ${\bf P}$ is the
closure of the image of $\sigma_{H,{\bf P},K}$ under the orthogonal
projection $\pi_\kappa$
along the $\kappa$-axis onto the $\eta$-${\bf k}$-plane. As
$\sigma_{H,{\bf P},K}\subset X$ and as $\pi_\kappa(X)$ is closed,
it follows that $\sigma_{H,{\bf P}}\subset \pi_\kappa(X)$.
Since $X$ contains the $\kappa$-axis, it now follows that
a point $(\eta,{\bf\impuls})$ can be in $\sigma_{H,{\bf P}}$ only if
$\{(\eta,{\bf k})\}\times\reals\subset X$. It follows that
Ineqs. (\ref{constraint1}) and (\ref{constraint2}) hold for all
$\kappa>0$ and all $\kappa<0$, respectively, and one finds
$-\E|{\bf P}|\leq H\leq\E|{\bf P}|$. But as the joint spectrum of
$H$ and ${\bf P}$ is a subspace of $\pi_\kappa(X)$, this
inequality, together with Lemma \ref{linear},
entails that $H$ is a linear function of ${\bf P}$, as stated.

It remains to prove Alternative (ii) for the case that
$H$ is not a linear function of ${\bf P}$.
By what we just proved, $X$ does not contain the $\kappa$-axis in this case,
so $K$ is a linear function of $H$ and ${\bf P}$ (cf. Lemma \ref{linear}),
i.e., there are $\beta\in\reals$ and ${\bf v}\in\reals^s$ such that
\begin{equation}\label{KundHundP}
K=-\beta H+{\bf v}{\bf P}.
\end{equation}
The vector ${\bf v}$ is unique up to a component that is perpendicular
to the smallest linear subspace $Y$ of $\reals^s$
containing the joint spectrum of the components of ${\bf P}$, so
${\bf v}$ can and will be chosen in $Y$.

If ${\bf vP}=0$, then
$K=-\beta H$, and Ineq. (\ref{constraint1}) reads
$$-\frac{\kappa}{\beta}\leq\E|{\bf k}|\frac{e^\kappa+1}{e^\kappa-1}.$$
for all $\kappa>0$ and all ${\bf k}\in Y$, so
$\beta>0$, which yields Alternative (ii).

In the remaining case, ${\bf vP}\not=0$, and since ${\bf v}\in Y$,
the unit vector ${\bf e_v}:=|{\bf v}|^{-1}{\bf v}$ is in $Y$.

If $\beta\not=0$, then Eq. (\ref{KundHundP}) and
the assumption that $H$ is not a function of ${\bf P}$ entail $K\not=0$,
so for each $\kappa>0$ and each $\lambda>0$,
one has $(\eta(\lambda,\kappa),\lambda{\bf e_v},\kappa)\in X$,
where
$$\eta(\lambda,\kappa)
:=-\frac{1}{\beta}(\kappa+\lambda {\bf e_v}{\bf v})
 =-\frac{1}{\beta}(\kappa+\lambda|{\bf v}|).$$
Since $X\subset\sigma^{(\E)}$, Ineq. (\ref{constraint1}) yields
$$-\frac{1}{\beta}(\kappa+
\lambda|{\bf v}|)\leq\lambda\E\frac{e^\kappa+1}{e^\kappa-1}$$
for all $\kappa,\lambda>0$, so $\beta>0$ and $|\frac{{\bf v}}{\beta}|\leq\E$,
and putting ${\bf u}:={\bf v}{\beta}$, one obtains Alternative (ii).

We can now complete the proof by showing that in the remaining case 
that $\beta=0$, one arrives at $K=0$, so that 
$\go$ is a trace, i.e., a KMS-state at infinite temperature.
As $H$ is not a linear function of ${\bf P}$, while $\beta=0$ implies
$K={\bf vP}$, $H$ cannot be a linear function of $K$ and ${\bf P}$, so
$X$ must contain the $\eta$-axis. But if
$K$ did not equal zero, there would exist a ${\bf k}\in\reals^s$ such that
${\bf vk}>0$ and $(\eta,{\bf k},{\bf vk})\in X$
for all $\eta\in\reals$, so Ineq. (\ref{constraint1}) would imply
$\eta\leq\E|{\bf\impuls}|
\frac{e^{{\bf v}{\bf k}}+1}{e^{{\bf v}{\bf k}}-1}$
for all $\eta\in\reals$.
As ${\bf vk}>0$ by assumption, this is impossible, so $K=0$, as
stated.
\Halmos

\bigskip
The next proposition considers the case that $\go$ is not faithful.

\begin{proposition}\label{ifnot}
Let $\go$ be completely semipassive with efficiency bound
$\E$. If $\go$ is not faithful, then there exists a ${\bf u}\in\reals^s$
with $|{\bf u}|\leq\E$
such that $H+{\bf uP}\geq0$.
\end{proposition}
{\it Proof.} As $\go$ is cyclic with respect to $\dM$, it is separating with
respect to $\dM'$. As the
projection operator $E_\h$ onto the closed subspace
$\h:=\overline{\dM'\gO}$ is easily seen to be an element of $\dM$, the
algebra $E_\h\dM E_\h:=\{E_\h ME_\h:\,M\in\dM\}$ is a von Neumann
subalgebra of $\dM$, and with respect to the von Neumann algebra
$$\dN:=\{E_{\h}M|_\h:\,M\in\dM\}$$
of operators in the Hilbert space $\h$,
$\gO$ is both cyclic and separating.  It is also
straightforward to check that the representation $V$ maps $\h$ and
$\h^{\perp}$ onto themselves, that it strongly commutes with $E_\h$
and, hence, implements automorphisms of $\dN$.

Now let $\Delta$ be the modular operator of $\dN$ and $\gO$, and
define the positive operator
$\tilde\Delta:=\Delta E_\h$. One checks (cf. App. \ref{TTTheory}) that
$\tilde\Delta$ strongly commutes with $H$, so one can consider
the joint spectrum  $\sigma_{H,{\bf P},\tilde\Delta}$ of
$H$, ${\bf P}$, and $\tilde\Delta$.

If $A\in\dM$, then $B:=AE_\h$ lies in $\dM$
as well, and inserting $B$ into Ineq. (\ref{windmill}) yields,
after the procedure followed earlier,
$$-H(1-\tilde\Delta)e^{-H^2-|{\bf P}|^2}
\leq\E|{\bf P}|(1+\tilde\Delta)e^{-H^2-|{\bf P}|^2}.$$
It follows that
$\sigma_{H,{\bf P},\tilde\Delta}$ is a subset of the set
$$\sigma^{(\E)}:=\{(\eta,{\bf\impuls},\delta)
\in\reals\times\reals^s\times\reals^{\geq0}:\,
-\eta(1-\delta)\leq\E|{\bf\impuls}|(1+\delta)\}.$$
The points in $\sigma^{(\E)}$ of the form $(\eta,{\bf k},0)$
satisfy the estimate
\begin{equation}\label{constraint3}
-\eta\leq\E|{\bf\impuls}|.
\end{equation}
The spectrum $\sigma_{H,{\bf P},\tilde\Delta}$ contains at least one
such point: as $\gO$ is not separating with respect to $\dM$,
while being separating with respect to $\dN$ by construction, one has
$\dM\not=\dN$ and $\h\not=\H$, and as $\tilde\Delta$ annihilates all elements
of $\h^\perp\not=\{0\}$, the elements of $\h^\perp$ are zero eigenvectors
of $\tilde\Delta$. Meanwhile, the representation $V$ maps
the subspace $\h^\perp$ onto itself, so it follows that
$H|_{\h^\perp}$ and ${\bf P}|_{\h^\perp}$ are self-adjoint
operators in $\h^\perp$, whose spectral projections are
restrictions of the corresponding spectral projections of
$H$ and $P$, respectively.
But this implies that $\sigma_{H,{\bf P},\tilde\Delta}$
contains some point of the form $(\eta,{\bf\impuls},0)$.

Next we prove that
all points in $\sigma_{H,{\bf P},\tilde\Delta}$ must satisfy Ineq.
(\ref{constraint3}) (even though not all of them are
of the form $(\eta,{\bf k},0)$).

To show that the opposite case cannot occur, choose
$(\eta,{\bf k},0)\in\sigma_{H,{\bf P},\tilde\Delta}$, and
let $(\eta',{\bf\impuls}',\delta')$ be any point
in $\sigma_{H,{\bf P},\tilde\Delta}$ that violates
Ineq. (\ref{constraint3}). As $\go$ is completely
semipassive, it follows that
$$(\eta+n\eta',{\bf\impuls}+n{\bf\impuls}',0\cdot n\delta')\in\sigma^{(\E)}
\qquad\forall n\in\naturals,$$
so
$$-(\eta+n\eta')\leq\E|{\bf\impuls}+n{\bf\impuls}'|
\qquad\forall n\in\naturals.$$
Choosing $n$ sufficiently
large, one now arrives at a contradiction with
the assumption that $-\eta'>\E|{\bf\impuls}'|$. It follows
that Ineq. (\ref{constraint3}) must hold for all elements of
$\sigma_{H,{\bf P},\tilde\Delta}$, as stated.

If one now applies Lemma \ref{projection}, one finds Ineq.
(\ref{constraint3}) for all $(\eta,{\bf k})\in\sigma_{H,{\bf P}}$.
But by complete semipassivity, the corresponding estimate
should hold for all tensorial powers of $\go$, which, as above,
implies that $\sigma_{H,{\bf P}}$ is a subset of
a sub-semigroup of $\Rd$ whose elements satisfy Ineq.
(\ref{constraint3}). But such a semigroup must be a subset of a
half space whose elements satisfy Ineq. (\ref{constraint3}) as well,
so it follows from Lemma \ref{projection} that there exists
a ${\bf u}\in\reals^s$ such that $|{\bf u}|\leq\E$ and such that
the operator $H+{\bf uP}$ is positive, which is the statement.

\Halmos

\bigskip
Summing up our results, one now obtains the following theorem:

\begin{theorem}\label{Hauptsatz}
The state $\go$ is completely semipassive with efficiency bound
$\E$ if and only if there exists a ${\bf u}\in\reals^s$ with
$|{\bf u}|\leq\E$ such that with respect to $H+{\bf uP}$,
$\go$ is a ground state or a KMS-state at a finite
inverse temperature $\beta\geq0$.
\end{theorem}

In a relativistic theory,
the generator of the time evolution of the system moving at
velocity ${\bf u}<c$ is not $H+{\bf u}{\bf P}$, but
$\gamma(H+{\bf u} {\bf P})$,
where $\gamma=(1-|{\bf u}|^2/c^2)^{-\frac{1}{2}}\equiv
(1-|{\bf u}|^2)^{-\frac{1}{2}}$. Theorem
\ref{Hauptsatz} still holds without any modification, but the
inverse temperature of the system is not the
parameter $\beta$ found there, but
$\beta/\gamma$.

\section{Semipassivity and the chemical potential}\label{chemical potential}
Above, we have considered the operators ${\bf P}$ as the generators
of the spatial translations.
But that ${\bf P}$ plays this concrete role,
has not been used in the proofs, so
other applications can easily be thought of. A
diffusion of particles or charges from or into infinite reservoirs
can take place in a stationary fashion (cf. also \cite{Rue00,GL00,
AHKT,Haa92,Kas76}). In this case, there should be
a vector $(N_1,\dots,N_n)=:N$ of self-adjoint
operators that can modify the generator of the time evolution
accordingly. Again, we assume that $N_1\gO=\dots=N_n\gO=0$.

Such diffusion processes can make cycles perform work
(cf. also \cite{Nar82}), and the condition
of semipassivity with respect to $N$
means that any work performed by a cycle
can be performed by these effects only. The state $\go$ is
semipassive with respect to $N$ if there exists a nonnegative constant
$\E_N$ such that
$$-\langle W\gO,HW\gO\rangle\leq\langle W\gO,\E_N|N|W\gO\rangle$$
for all $W\in U_1(\dM)$ with $[H,W]\in\dM$ and $[N,W]\in\dM$.
Mimicking the proofs of Props. \ref{rkms theorem} and \ref{ifnot},
one directly obtains the following corollary.
\begin{corollary}
With respect to $N$,
the state $\go$ is completely semipassive with efficiency bound
$\E_N$ if and only if there is a vector
$\mu:=(\mu_1,\dots,\mu_n)$ such that with respect to $H+\mu N$,
$\go$ is either a ground state or
a KMS-state at a finite inverse temperature
$\beta\geq0$.
\end{corollary}
The vector $\mu$ collects the chemical potentials associated with
the different particles or charges.

\section{Passivity and vacuum states}\label{vacuum}

A cycle should perform work only if there is either some flow of matter
or if the cycle is driven by a nonstationary inertial force due to the
observer's motion.
Since matter is completely absent in vacuum, each
vacuum state should be passive in every uniformly
accelerated frame (whose acceleration may be zero).

In this section, we show that if a pure state $\go$ behaves this way,
then it is invariant under spacetime translations, and the
four-momentum spectrum is contained in the forward lightcone
(spectrum condition); these are the familiar defining
properties of a vacuum state.
For the case that, in addition, $\dM$ arises from a
relativistic quantum field theory and the vacuum state exhibits passivity
in each uniformly accelerating frame, it will be shown in the next section
that in the eyes of each uniformly accelerating observer,
$\go$ is a KMS-state at
a nonzero and finite positive temperature proportional to the
acceleration, which is the Unruh effect.

As above, let $\go$ be a state of a von Neumann algebra $\dM$, let
$\go$ be induced by a cyclic vector $\gO$ as above, and assume
that there is a strongly continuous
unitary representation $V$ of $(\Rd,+)$ with generators $H$ and
${\bf P}$ and with the property that
$V(x)\dM V(x)^*=\dM$ for all $x\in\Rd$.

If $\go$ is a vacuum state, then the above considerations
suggest that it should, in particular, be passive with respect to
each Hamiltonian of the form
$\gamma(H+{\bf v}{\bf P})$ with
$|{\bf v}|<c=1$, and
$\gamma=(1-{\bf v}^2/c^2)^{-1/2}$.
As passivity implies stationarity, it follows that
$\go$ should be invariant under all spacetime translations;
again, we can assume without loss that $H\gO=P_1\gO=\dots=P_s\gO=0$.

If $V(x)\in\dM$ for all $x\in\Rd$, which holds, in particular, if $\go$
is a pure state, as $\dM=\dB(\H)$ in this case, one
can prove the following:

\begin{proposition}\label{positive}
Let the state $\go$ exhibit passivity with respect to
$\gamma(H+{\bf v}{\bf P})$ for each ${\bf v}\in\reals^s$ with
$|{\bf v}|<1$, and suppose that $V(x)\in\dM$ for all $x\in\Rd$. Then
the joint spectrum of $H$ and ${\bf P}$ is contained in the cone
$$V_+:=\{(\eta,{\bf k})\in\Rd:\,\eta\geq0,\eta^2-{\bf k}^2\geq0\},$$
i.e., the spectrum condition holds.
\end{proposition}

{\it Proof.}
As $\go$ is passive with respect to $K:=\gamma(H+{\bf vP})$ for each
${\bf v}\in\reals^s$ with $|{\bf v}|<1$, it follows
(see Ineq. (3.8) in \cite{PW78}) that
\begin{equation}\label{passivity implies}
\langle A\gO,Ke^{-K^2}A\gO\rangle+\langle A^*\gO,Ke^{-K^2}A^*\gO\rangle
\quad\forall A\in\dM.
\end{equation}
Note that $Ke^{-K^2}$ is a bounded operator.

As $V(x)\in\dM$ for all $x\in\Rd$, the spectral
projection $E:=E_K(\{0\})$ of
$K$ associated with $\{0\}$ is an element of $\dM$
for every ${\bf v}\in\reals^s$, and
$A:=(1-E)BE\in\dM$ for all $B\in\dM$.
Inserting $A$ into Ineq. (\ref{passivity
implies}), and taking into account that $K\gO=0$, one finds
\begin{equation}\label{passivity implies more}
\langle B\gO,Ke^{-K^2}(1-E)B\gO\rangle+\underbrace{\langle E B^*(1-E)\gO,
Ke^{-K^2}B^*(1-E)\gO\rangle}_{=0}\geq0
\end{equation}
(note that $E\gO=\gO$ by construction),
so $Ke^{-K^2}(1-E)$ is a positive (bounded)
operator.
Since the function $x\mapsto xe^{-x^2}$ preserves signs,
it follows that $K(1-E)$ is a positive operator as well.
But on the other hand, $KE=0$, so $K$ is positive.
This immediately implies the spectrum condition;
note that the forward light cone is an intersection of
half spaces.

\Halmos

If, conversely, $\go$ is known to be
spacetime translation invariant and to satisfy
the spectrum condition,
it can be shown that the unitary operators
$V(x)$, $x\in\Rd$, are elements of $\dM$
(\cite{Ara64}, see also Thm. III.3.2.4 in \cite{Haa92}),
and $\go$ can be decomposed into pure states
that are invariant under spacetime translations and
satisfy the spectrum condition as well
(see, e.g., Sect. III.3.2 in \cite{Haa92}).

\section{Passivity and the Unruh effect} \label{BW-theorem}

Let $\dM$, $\go$, $V$ and $\gO$ be as above.
We now need some basic structures
of local quantum fields, which associate von Neumann algebras
$\dM(\O)$ of local observables with all bounded open spacetime regions
$\O\subset\Rd$ in such a way that the following conditions are satisfied:
\begin{quote}
{\bf (A) Isotony.} If $\O$ and $P$ are bounded open regions in $\Rd$
such that $\O\subset P$, then $\dM(\O)\subset\dM(P)$.

{\bf (B) Locality.} If $\O$ and $P$ are spacelike separated bounded open
regions in $\Rd$ and if $A\in\dM(\O)$ and $B\in\dM(P)$, then $AB=BA$.

{\bf (C) Spacetime Translation Covariance.} The
representation $V$ of $(\Rd,+)$ satisfies
$$V(x)\dM(\O)V(x)^*=\dM(\O+x)$$
for all bounded open sets $\O\subset\Rd$ and for all $x\in\Rd$.

{\bf (D) Spectrum Condition.} The joint spectrum of the generators of
$V$ is contained in the closed forward light cone.
\end{quote}
$\dM$ is assumed to be the smallest
von Neumann algebra that contains all local algebras $\dM(\O)$
associated with bounded open regions.

The trajectory of a (pointlike) observer who is uniformly accelerated
in the 1-direction with acceleration $a$ can be translated to the
curve
$$t\mapsto\frac{c^2}{a}\left(\sinh\frac{at}{c},
\cosh\frac{at}{c},0,\dots,0\right),\qquad\tau\in\reals,$$
where $t\in\reals$
denotes the accelerated observer's eigentime. The wedge
$W_1:=\{x\in\Rd:\,x_1>|x_0|\}$, which is referred to as the {\em
Rindler wedge}, is the region of all spacetime points the accelerated
observer can communicate with using causal signals. Therefore, the
elements of the algebra $\dM(W_1)$ are precisely those observables the
uniformly accelerated observer can measure. The images of $W_1$ under
Poincar\'e transformations are referred to as {\em wedges}.

We assume that some uniformly accelerated observer exists:
\begin{quote}
{\bf (E)} There is a self-adjoint operator
$K_1$ generating, within $W_1$, the free dynamics of the uniformly
accelerating observer, i.e., $$e^{i\tau K_1}\dM(\O)e^{-i\tau
K_1}=\dM(\Lambda_1(\mbox{$\frac{a}{c}$}\tau)\O)$$ for all $\tau\in\reals$ and
all bounded open sets $\O\subset W_1$.
$K_1$ strongly commutes with $P_2,\dots,P_s$, and $K_1\gO=0$.
\end{quote}
Here, $\Lambda_1(\frac{a}{c}\tau)$ denotes the Lorentz boost by
$\frac{a}{c}\tau$ in the 1-direction.
$\dM$ is not yet assumed to be
covariant under a full representation of the Poincar\'e group,
although the assumption that $K_1$ strongly commutes with
$P_2,\dots,P_s$ is already a part of this condition.

\begin{proposition}%\label{Unruh effect}
With the above assumptions, assume $\go$ to exhibit passivity with respect
to the dynamics generated by $K_1$. Then $\go$
is a KMS-state of $\dM(W_1)$ with respect to $K_1$ at
the Unruh temperature
$$T_U=\frac{\hbar a}{2\pi ck}.$$
\end{proposition}

{\it Proof.} As a consequence of the spectrum
condition, $\gO$ is cyclic not only with respect to $\dM$, but even
with respect to $\dM(W_1)$ and $\dM(-W_1)$ (cf. \cite{Buc75}, p. 279).
By Prop. 2.2 in \cite{BF82}, it also
follows from the spectrum condition and locality that the space of
vectors that are invariant under translations in the 2-direction is
1-dimensional (and, thus, spanned by $\gO$), and as $\gO$ is cyclic
with respect to $\dM(W_1)$,
the state $\go|_{\dM(W_1)}$ weakly clusters (in the sense of \cite{PW78})
with respect to the translations in the 2-direction,
(note that $W_1$ is invariant under these translations).
As the translations in the 2-direction strongly commute with $K_1$, one
can apply Thm. 1.3 in \cite{PW78} to conclude that $\go$ must be a
KMS-state or a ground state with respect to the dynamics generated by
$K_1$. In particular, $\go$ exhibits complete passivity.

As $\gO$ is cyclic with respect to $\dM(-W_1)$ and as $-W_1$
is spacelike with respect to $W_1$, locality implies that
$\gO$ is separating with respect to $\dM(W_1)$, so $\go|_{\dM(W_1)}$
is faithful.
It follows that $\go|_{\dM(W_1)}$ is a KMS-state at some inverse temperature
$0\leq\beta<\infty$ (use, e.g., Prop. \ref{rkms theorem}
above for $\E=0$). As the action of the
modular unitary operators is nontrivial, their generator differs from
zero, so $\go|_{\dM(W_1)}$ cannot be a trace, and one even has
$\beta>0$. The modular group is generated by the operator
$\beta\frac{\hbar a}{c}K_1$.

It is now easy to compute $\beta$ and
the {\em Unruh temperature} $T_U:=(k\beta)^{-1}$. It follows from Thm.
II.9 in \cite{Bor92}\footnote{
See also \cite{Flo98} for a considerably shorter proof.}
that the spectrum condition entails
$$\exp\left(it\beta\frac{\hbar a}{c}K_1\right)V(x)
\exp\left(-it\beta\frac{\hbar a}{c}K_1\right)=
V(\Lambda_1(-2\pi t)x)$$
for all $x\in\Rd$,
so it is evident that $\beta\frac{\hbar a}{c}=2\pi$, whence
the stated formula for the Unruh temperature follows.

\Halmos

Assuming the statement of this proposition in all Lorentz frames,
group cohomological arguments imply that
$V$ and the operators $K_W$ associated with all wedges $W$,
generate a representation
of the proper Poincar\'e group $\P_+$ \cite{BGL93}, and
this representation is even a representation
of the restricted Poincar\'e group $\P_+^\uparrow$ \cite{GL95}. The
modular conjugation of the Rindler wedge implements a
\pct-symmetry, i.e., a spatial reflection of the 1-component, a
time reflection, and a charge conjugation \cite{GL95}.
This fact was found to imply the spin-statistics for massive
(para-) bosonic and (para-) fermionic particles
\cite{GL95,Kuc95}, and as these proofs did not use any
spinor calculus, it was possible to generalize them to
conformal quantum field theories \cite{GL96}, to
massive particles with braid group statistics in 1+2 dimensions
such as anyons \cite{Lon97,Mun98}, and to special quantum field theories
on (sufficiently symmetric) curved spacetimes \cite{GLRV}.

%{\it (ii) Is the Unruh effect classical or quantum?} The Unruh effect is a true
%quantum effect ($T_U\to0$ as $\hbar\to0$) and a
%truly relativistic effect ($T_U\to0$ as
%$c\to\infty$). Although the above characterization of a
%vacuum already makes sense in a
%classical setting, a nonzero Unruh temperature can
%only be obtained in a relativistic quantum theory.
%A similar situation is met in Unruh's
%original proof (cf. the discussion in \cite{PaV99}).

%{\it (iii) Nonstationary inertial forces destroy thermodynamic equilibrium.}
%Instead of assuming that $K_1$ generates the dynamics of a
%uniformly accelerated observer and that $\go$ is passive with respect
%to this dynamics, one may also start from the modular group of
%$\dM(W_1)$ and require that
%this group acts geometrically on the net in some sense,
%which evidently is necessary if it is to give rise to any time
%evolution. If this is assumed, then the modular group of the Rindler wedge
%must necessarily implement the Lorentz boosts that leave the wedge
%invariant (\cite{Kuc00}, cf. also the related results
%reviewed in \cite{Bor00}). This shows, in particular, that
%nonstationary inertial forces necessarily destroy thermodynamic equilibrium not merely
%in classical thermodynamics, but also in the setting of local quantum
%fields, which provides further justification of the above characterization
%of the vacuum.

\section{Conclusion}\label{conclusion}

The behaviour of cycles can be used to characterize thermodynamic
equilibrium states in a covariant fashion.  Cycles
cannot extract any energy from a
system in thermodynamic equilibrium
by performing exterior work, i.e., thermodynamic equilibrium states
exhibit passivity. It
follows that if a thermodynamic equilibrium state is observed from a
uniformly moving frame of reference, it ceases to be a thermodynamic
equilibrium state, as cycles can perform work there. The amount of work
a cycle can perform when applied to a moving thermodynamic equilibrium
state is bounded by the amount of work an ideal windmill or turbine
could perform; this property is called {\em semipassivity}, and the
factor $\E\geq0$ characterizing the bound is called an {\em efficiency
bound}. The Zeroth Law justifies a strengthening of passivity and
semipassivity called {\em complete passivity} and {\em complete
semipassivity}, respectively.

For the description of homogeneous states, the condition of complete
semipassivity turns out to be the appropriate generalization of
complete passivity to moving frames of reference. If it holds, an
inertial frame can be found where the system is in thermodynamic
equilibrium. Semipassivity can also be used to measure the violation
of passivity due to stationary diffusion processes and to define
the corresponding chemical potentials.

When applied to states without matter, cycles should
not perform any work unless there are nonstationary
inertial forces to drive them. Each pure state behaving
this way satisfies the spectrum condition, and in the general
setting of local quantum field theory in Minkowski spacetime,
such a state appears as a thermodynamic equilibrium state at the
Unruh temperature $\frac{\hbar a}{2\pi ck}$ to each
observer who is uniformly accelerated with acceleration $a$.

\subsection*{Acknowledgements}
The author thanks D. Arlt, D. Buchholz, K. Fredenhagen, N. P.
Landsman, J. Mund, and R. Verch for critically reading
preliminary versions of the manuscript.

This work has been supported by a Feodor-Lynen grant of the
Alexander-von-Humboldt-Stiftung, a Casimir-Ziegler grant of
the Nordrhein-Westf\"ali\-sche Akademie der Wissenschaften, and
by the Stichting Fundamenteel Onderzoek der Materie.

\begin{appendix}
\section*{Appendix}

\section{Some Tomita-Takesaki theory}\label{TTTheory}
The modular theory
founded by Tomita and Takesaki \cite{Tak} plays an important
role in quantum field theory and quantum statistical mechanics (cf., e.g.,
\cite{Haa92,Bor00}). It is used in the above proofs, so
for the convenience of the reader, some
relevant facts and notation of Tomita-Takesaki theory are
summarized here in a nutshell.

As above, let a state $\go$ of $\dM$ be induced by the cyclic vector
$\gO$. Suppose that $\go$ is faithful, then $\gO$ is also  {\em
separating} with respect to $\dM$, i.e., given an $A\in\dM$, $A\gO=0$
implies $A=0$. This implies that the map $$A\gO\mapsto A^*\gO,\qquad
A\in\dM,$$ defines an antilinear operator on the dense domain
$\dM\gO$. This operator is closable, and its closure $S$ can, like a
complex number, be polar-decomposed, $S=J\Delta^{1/2}$, into a
positive operator $\Delta^{1/2}$ (its ``modulus'') and an antiunitary
operator $J$ (its ``phase''). $\Delta$ and $J$, like $S$, leave $\gO$
fixed by construction, and $J$ is a conjugation, i.e., $J^2=1$. By a
theorem of Tomita and Takesaki \cite{Tak}, one has
\begin{align*}
\Delta^{it}\dM\Delta^{-it}&=\dM\qquad\forall t\in\reals;\\
J\dM J&=\dM'.
\end{align*}
With respect to
the dynamics $A\mapsto\Delta^{it}A\Delta^{-it}=:A_t$, $A\in\dM$, $t\in\reals$,
the state $\go$ satisfies the KMS-condition:
\begin{align*}
\langle\gO,AB_t\gO\rangle&=\langle A^*\gO,\Delta^{it}B\Delta^{-it}\gO\rangle
=\langle J\modop A\gO,J\modop\Delta^{it}B^*\Delta^{-it}\gO\rangle\\
&=\langle \modop\modopt B^*\modopmt,\modop A\gO\rangle
=\langle\gO,\modopt B\modopmt\Delta A\gO\rangle\\
&=\langle\Delta\gO,\modopt B\modopmt\Delta A\gO\rangle
=\langle\gO,B_{t-i}A\gO\rangle
\end{align*}
(cf. also Lemma 8.1.10 (p. 351) in \cite{LiB92}),
and for each faithful state $\go$ of a
von Neumann algebra there is only one strongly continuous
automorphism group of $\dM$ with this property \cite{Tak}.
The positive operator
$\Delta$ is referred to as the {\em modular operator of $\dM$ and $\gO$},
the group of the automorphisms $A\mapsto A_t$,
$t\in\reals$, of $\dM$ is called the {\em modular (automorphism) group}, and
the conjugation $J$ is the {\em modular conjugation} of $\dM$ and $\gO$.

If $U$ is a unitary operator in $\H$ such that $U\dM U^*=\dM$ and
$U\gO=\gO$, then one has, for all $A\in\dM$:
$$USU^*A\gO=US(U^*AU)\gO=U(U^*AU)^*\gO=A^*\gO=SA\gO,$$
and this suffices to conclude that
$USU^*=S$. The fact that
$$J\Delta^{1/2}=S=USU^*=UJ\Delta^{1/2}U^*=UJU^*U\Delta^{1/2}U^*,$$
together with the uniqueness of the polar decomposition of a closed linear
or antilinear operator, implies that $UJU^*=J$, that
$U\Delta^{1/2}U^*=\Delta^{1/2}$, and that $U\Delta^{it}U^*=\Delta^{it}$.
It follows that each self-adjoint operator $G$ in $\H$ with $G\gO=0$
and with $e^{itG}\dM e^{-itG}=\dM$ for all $t\in\reals$,
strongly commutes with $K$ and $J$. This implies $e^{itG}=Je^{itG}J$
for all $t\in\reals$, and it is not difficult to conclude
that $iG=J(iG)J=-iJGJ$, so one arrives at the relation $JGJ=-G$, which is
used for $G=H$, $G=K$, and $G={\bf P}$ in the text.

\section{Some remarks on joint spectra}\label{jspec}

For the reader's convenience, we recall some basic facts on joint spectra
of strongly commuting self-adjoint operators. We work with three operators
for notational convenience; the generalization to $n$ operators is
straightforward.

Let $A$, $B$, and $C$ be self-adjoint operators that commute strongly,
i.e., whose spectral projections commute and, hence, define a
product spectral measure. The {\it joint spectrum} $\sigma_{A,B,C}$
of $A$, $B$, and $C$ is the support of the product measure $E_{A,B,C}$
of the spectral measures $E_A$, $E_B$, and $E_C$. It is a closed set by
construction. A point $(x,y,z)\in\reals^3$
is in $\sigma_{A,B,C}$ if and only if for all $\eps>0$, one has
$$E_A([x-\eps,x+\eps])E_B([y-\eps,y+\eps])E_C([z-\eps,z+\eps])\not=0.$$
The following lemmas are used in the above proofs.
\begin{lemma}\label{linear}
Let $X$ be a two-dimensional subspace of $\reals^3$ that does not
contain the $z$-axis. If $\sigma_{A,B,C}\subset X$, then $C$ is a
linear function of the operators $A$ and $B$.
\end{lemma}
{\it Proof.}
As $X$ does not contain the $z$-axis, there is a
linear function $f:\reals^2\to\reals$ such that
$X=\{(x,y,f(x,y)):\,(x,y)\in\reals^2\}$, and if
$I\subset\reals$ is any Borel set, one checks that
\begin{align*}
E_C(I)&=E_A(\reals)E_B(\reals)E_C(I)\\
&=E_{A,B,C}(\reals^2\times I)=E_{A,B,C}((\reals^2\times I)\cap X)\\
&=E_{A,B,C}((f^{-1}(I)\times\reals)\cap X)=E_{A,B,C}((f^{-1}(I)\times\reals)\\
&=E_{A,B}(f^{-1}(I))E_C(\reals)=E_{A,B}(f^{-1}(I)).
\end{align*}
\Halmos

\begin{lemma}\label{projection}
Let $\pi_z$ denote the orthogonal projection along the z-axis onto
the x-y-plane. Then
$$\sigma_{A,B}=\overline{\pi_z(\sigma_{A,B,C})}.$$
\end{lemma}
{\it Proof.} If $(x,y,z)\in\sigma_{A,B,C}$, then for each $\eps>0$,
one has
\begin{align*}
F_\eps&:=E_A([x-\eps,x+\eps])E_B([y-\eps,y+\eps])\\
&=F_\eps E_C(\reals)\geq F_\eps E_C([z-\eps,z+\eps])\not=0,
\end{align*}
so $(x,y)\in\sigma_{A,B}$, proving
$\pi_z(\sigma_{A,B,C})\subset\sigma_{A,B}$. As $\sigma_{A,B}$ is closed,
it follows that $\overline{\pi_z(\sigma_{A,B,C})}\subset\sigma_{A,B}$.

If, conversely, $(x,y)\in\sigma_{A,B}$, then one has
$F_\eps\not=0$ for all $\eps>0$, so if the set
$$M_\eps:=\pi_z^{-1}([x-\eps,x+\eps]\times[y-\eps,y+\eps])$$
had empty intersection with $\sigma_{A,B,C}$,
the product of $F_\eps$ with all spectral projections of $C$
would equal zero, and in particular, $F_\eps=F_\eps E_C(\reals)=0$,
which is in conclict with $(x,y)\in\sigma_{A,B}$.

We conclude that each open neighbourhood of $(x,y)$ contains a point in
$\pi_z(\sigma_{A,B,C})$, so
$\sigma_{A,B}\subset\overline{\pi_z(\sigma_{A,B,C})}$, and the
proof is complete.

\Halmos

\end{appendix}

\end{document}